\documentclass[aps,physrev,preprint,groupedaddress,showkeys,superscriptaddress]{revtex4-2}

\usepackage{graphicx}
\usepackage{amssymb}
\usepackage{amsmath}
\usepackage[hidelinks]{hyperref}

\newcommand\lp{\left(}
\newcommand\rp{\right)}
\newcommand\ls{\left[}
\newcommand\rs{\right]}
\newcommand\lbr{\left\lbrace}
\newcommand\rbr{\right\rbrace}
\newcommand\e{\mathrm{e}}

\usepackage{xcolor}

\begin{document}
	
	
	\title{Offer of a reward does not always promote trust in spatial games}
	
	
	\author{Haidong Zhang}
	\affiliation{College of Systems Engineering, National University of Defense Technology, Changsha~410073, China}
	
	\author{Chaoqian Wang}
	\email[Contact author: ]{CqWang814921147@outlook.com}
	\affiliation{School of Mathematics and Statistics, Nanjing University of Science and Technology, Nanjing 210094, China}
	
	\author{Shuo Liu}
	\affiliation{College of Systems Engineering, National University of Defense Technology, Changsha~410073, China}
	
	\author{Charo~I.~del~Genio}
	\affiliation{School of Mathematics, North University of China, Taiyuan~030051, China}
	\affiliation{Institute of Interdisciplinary Intelligent Science, Ningbo University of Technology, Ningbo~315211, China}
	\affiliation{Institute of Smart Agriculture for Safe and Functional Foods and Supplements, Trakia University, Stara Zagora~6000, Bulgaria}
	
	\author{Stefano Boccaletti}
	\affiliation{School of Mathematics, North University of China,	Taiyuan~030051, China}
	\affiliation{CNR, Institute for Complex Systems, Florence~50019, Italy}
	\affiliation{Research Institute of Interdisciplinary Intelligent Science, Ningbo University of Technology, Ningbo~315104, China}
	
	\author{Xin Lu}
	\email[Contact author: ]{xin.lu.lab@outlook.com}
	\affiliation{College of Systems Engineering,	National University of Defense Technology, Changsha~410073, China}
	
	
	\date{\today}
	
	\begin{abstract}
		Trust is one of the cornerstones of human society. One of the evolutionary
		pressure mechanisms that may have led to its emergence is the presence of
		incentives for trustworthy behavior. However, this type of reward has received
		relatively little attention in the context of spatial trust games, which are
		often used to build models in evolutionary game theory. To fill this gap,
		we introduce an inter-role reward mechanism in the spatial trust game, so
		that an investing trustor can choose to pay an extra cost to reward a trustworthy
		trustee. With extensive numerical simulations, we find that this type of reward
		does not always promote trust. Rather, while moderate rewards break the dominance
		of mistrust, thereby favoring investment, excessive rewards eventually stimulate
		a nonreturn strategy, ultimately suppressing the evolution of trust. Additionally,
		lower reward costs do not necessarily promote trust. Instead, more costly,
		but not excessive, rewards enhance the advantage of the original investment,
		consolidating the clusters of rewarders and improving trust. Our model thus
		provides evidence about the counterintuitive nature of the relationship between
		trust and rewards in a complex society.
	\end{abstract}
	
	
	\maketitle
	
	\section{Introduction}\label{secintro}
	
	Evolutionary game theory builds on classical
	game theory by adding mechanisms like selection, mutation, and
	imitation, to model the emergence and stabilization of cooperation
	and other complex behavioral strategies within groups~\cite{nowak1992evolutionary,traulsen2009stochastic,cressman2014replicator}.
	The explicit incorporation of spatial structures in the models
	used, with agents interacting on complex networks, allowed researchers
	to investigate effects that cannot be captured by a mean-field,
	all-to-all description~\cite{hauert2004spatial,labyrnet,rong2016proper}. A key insight is that network topology may promote the emergence of cooperation, governed by a simple rule: Cooperation is favored when the benefit-to-cost ratio exceeds the network's average degree~\cite{ohtsuki2006simple}. Building on this rule, recent studies have established general conditions for cooperation across a wide range of topologies, from heterogeneous networks to systems with higher-order interactions~\cite{allen2017evolutionary, su2022evolution_PNAS, su2022evolution_NHB, sheng2024strategy}. Furthermore, advances in temporal network theory have shown that time-varying interactions and dynamically evolving networks can fundamentally alter evolutionary trajectories compared to static settings~\cite{li2020evolution, su2023strategy}. Despite these developments, lattice networks remain a fundamental benchmark for sustaining cooperation and give rise to rich emergent phenomena such as fixation, coexistence, and cyclic dynamics~\cite{nowak1992evolutionary,szabo2007evolutionary,perc2017evolutionary,szolnoki2014cyclic,rantala2023spatial,wang2024evolutionary}. These theoretical insights have transcended biology, economics, and social sciences, increasingly inspiring algorithmic innovations in multiagent optimization, swarm intelligence, and evolutionary computation~\cite{lucas2006evolutionary,duan2014swarm,meng2024dynamics,meng2025promoting,li2013evolutionary}.
	
	A paradigmatic model used to study interpersonal reciprocity
	in mathematical and computational social sciences is the trust
	game~\cite{berg1995trust}. In its classic form, this asymmetric
	game is played between a trustor and a trustee, with the trustor
	sending a portion of an endowment to the trustee, and the trustee
	subsequently deciding how much to return to the trustor. Under
	purely selfish preferences, this setup creates a social paradox:
	The trustee's payoff-maximizing response is to keep all the transfer,
	but in practice trust is common and can easily diffuse across
	interactions~\cite{cohen2021trust}. The study of evolutionary games on networks explains this apparent contradiction. In fact,
	while exploitation commonly arises in well-mixed populations,
	introducing a nontrivial spatial structure can instead promote
	cooperation~\cite{iranzo2011empathy,chica2018evolutionary,wang2024evolution,liu2024evolutionary,shihui2025evolutionary}.
	To model this, existing studies have focused on mechanisms such
	as reinforcement learning~\cite{ren2023reputation,zhang2024emergence},
	belief-based preferences~\cite{attanasi2025disclosure}, adaptive
	reputation systems~\cite{hu2021adaptive,tang2014reputation}, second-order reputation~\cite{li2022n}, and trust dynamics~\cite{han2021or}. However, the classic formulation
	of the trust game has one more main limitation, namely that, because
	of its asymmetric nature, it requires players to imitate strategies
	solely within the same roles~\cite{masuda2012coevolution,masuda2014evolution,lim2024truster,wang2024evolution,liu2024heterogeneously}.
	Recognizing this limitation, some scholars have considered role
	alternation, so that each player occupies each of the two roles
	half of the time~\cite{nowak2000fairness,hofbauer2003evolutionary,sigmund2010calculus}.
	However, this approach does not fully capture settings in which
	the roles are fixed, but agents in one role can evolve their strategy
	using information originating from the agents in the other role.
	
	\begin{figure*}[t]
		\centering
		\includegraphics[width=\textwidth]{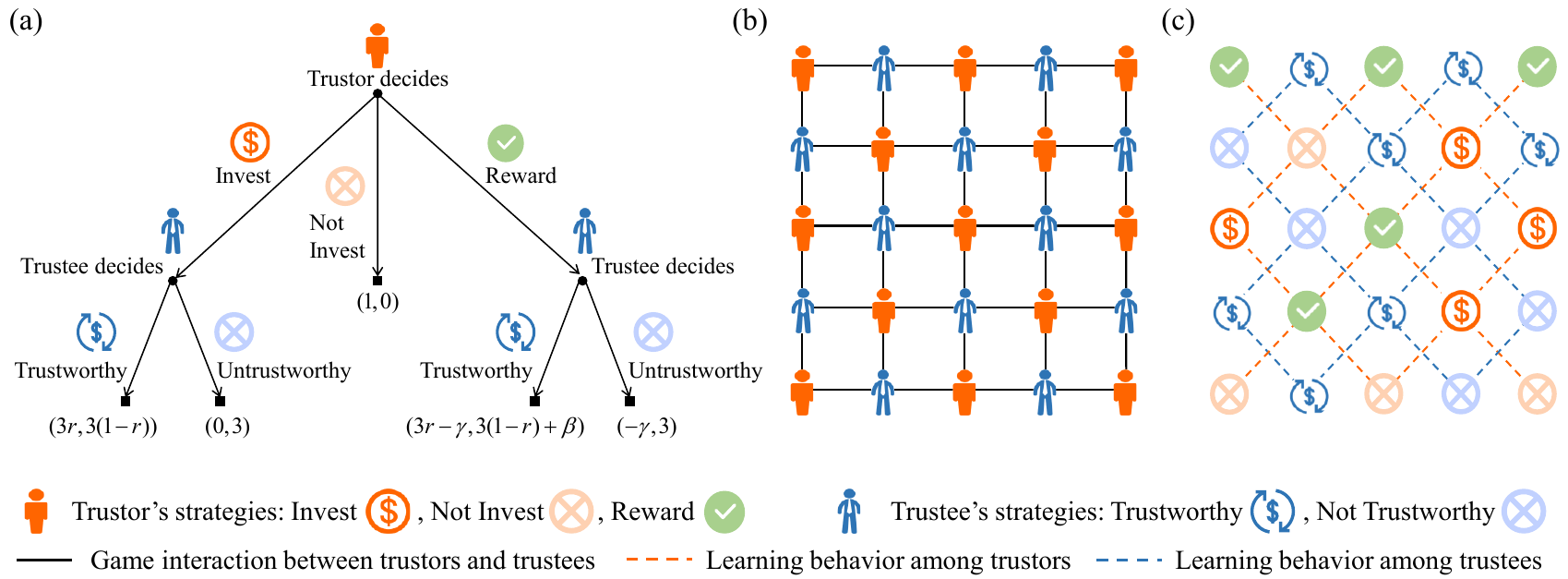}
		\caption{Trust game model with rewards and spatial interaction structure. (a) The game tree for the trust game with a reward mechanism. The trustor first decides	whether to invest~(I), not~invest~(N), or reward~(R). Then, the trustee decides whether
			to return~(T) or not~return~(U), and a single round of the game ends after these two
			stages. (b) Individuals interact with their nearest neighbors, calculating the average
			payoff from interactions with multiple neighbors. (c) Individuals engage in strategy
			learning with their second nearest neighbors, ensuring that strategies are updated
			among individuals in the same role.}\label{fig:network}
	\end{figure*}
	To bridge the two roles, the idea of reward has been extensively
	explored, in theoretical models as well as in behavioral experiments,
	with the overall result that third-party rewards effectively promote
	trust and cooperation~\cite{bolle1998rewarding,fiedler2017effect,ohtsubo2018within,fang2021evolutionary}.
	In fact, rewards function more effectively than punishments, and
	despite the associated costs and diminishing returns, because they
	entice players to remain in the game and to adopt cooperative strategies,
	particularly when multiple cooperative options are concurrently
	available~\cite{szolnoki2013rewarding,duong2021cost,han2022institutional,xiao2023evolution},
	and even though they may generate a ``free-rider aversion'' effect
	that can occasionally result in the reduction of individual contributions~\cite{alventosa2023pool}.
	However, although reward mechanisms have been extensively studied,
	a fundamental question remains, namely, how costly third-party rewards
	affect trust dynamics in structured populations with fixed roles.
	
	In this article, we fill this knowledge gap by investigating how inter-role rewards influence trust evolution in a spatial
	two-role trust game on a lattice network. Specifically, we extend
	the classical trust game by introducing a trustor-side reward
	strategy, where trustors can decide to bear some cost to reward
	trustworthy trustees. Our findings reveal a nonmonotonic, counterintuitive
	behavior of the system, so that rewards do not always promote
	trust and may even suppress it under certain conditions. Moreover,
	the effect of the cost of the reward incurred by the trustor
	is also opposite to what one would expect, as higher costs tend
	to improve trust, whereas lower ones do not necessarily promote it.

	\section{Model}\label{sec_model}
	\subsection{Trust game with reward}
	As described above, in the trust game, trustors decide
	at every time step whether to transfer a unitary endowment
	to trustees. Following the classic setup, if the trustor
	invests, the transferred amount is tripled upon reaching
	the trustee~\cite{berg1995trust,kumar2020evolution,wang2025inter}.
	The trustee then chooses between keeping all three units of
	wealth or returning a fraction~$r$ thereof to the trustor.
	For the trustor, investing is rational only when the trustee
	returns a fraction~$r$ such that $3r > 1$. Otherwise,
	keeping the initial endowment is more profitable than
	investing and receiving less than the invested amount
	in the best case, or nothing at all in the worst case.
	From the point of view of the trustee,
	if the trustor does not invest, they receive no payoff,
	making the question of their strategy moot. If instead
	the trustor invests, the trustee receives~3 by keeping
	everything, but only~$3\lp 1-r\rp$ by returning a fraction~$r$.
	This shows that the trustor has no dominant strategy,
	while the trustee's dominant strategy is to keep the entire
	amount.
	
	To increase the likelihood of trustors receiving some return,
	a third strategy can be introduced, namely that of trustors rewarding
	trustees who choose to return some of the invested wealth. Specifically,
	the reward strategy requires trustors to incur an additional
	cost $\gamma<1$ when investing. This cost is converted into a
	reward~$\beta$ that the trustees receive if they return the investment.
	Thus, this strategy functions as an incentive that fosters reciprocal
	behavior from the trustees. We set that reward cost $\gamma \in [0,1]$ and reward value $\beta \in [0.1, 100]$. The reason for setting cost $\gamma$ and reward benefit $\beta$ to different values and orders of magnitude stems from numerous real-world examples. For instance, a low-cost action may generate a disproportionately large benefit, as in the case where a simple word of encouragement yields a substantial payoff to the recipient, particularly under adverse conditions. Conversely, a high-cost action may produce little or even negative benefit, as when a costly gift fails to align with the recipient's preferences and is perceived unfavorably~\cite{szolnoki2010reward}.
	
	Summarizing, we can categorize the trustors' strategies
	into three types, namely, invest~(I), not~Invest~(N) and
	reward~(R), and the trustees' strategies into two types,
	namely, trustworthy~(T) and untrustworthy~(U), as illustrated
	in the game tree shown in Fig.~\ref{fig:network}(a). If
	the trustor chooses not to invest, the game step terminates,
	the trustor retains the initial capital of~1, and the trustee's
	payoff is~0. If the trustor opts to invest or to reward,
	the trustee then decides between~T and~U. All the possible
	payoffs are conveniently represented in the matrix
	\begin{equation}\label{eq:payoff_matrix}
		\begin{array}{c|cc}
			& \text{T} & \text{U} \\
			\hline
			\text{I} & \lp 3r, 3\lp1 - r\rp\rp & \lp 0, 3\rp \\
			\text{N} & \lp 1, 0\rp & \lp 1, 0\rp \\
			\text{R} & \lp 3r - \gamma, 3\lp 1-r\rp + \beta\rp & \lp -\gamma, 3\rp
		\end{array} \;
	\end{equation}
	Specifically, each pair ([trustor], [trustee]) in the payoff matrix corresponds to the respective payoffs obtained by an individual from the trustor population adopting the row strategy (I, N, or R) and an individual from the trustee population adopting the column strategy (T or U) when they interact.
	
	In the following, 
	$\Gamma = \lp \mathcal N, \lp S_i\rp_{i \in \mathcal{N}}, \lp u_i\rp_{i \in \mathcal{N}}\rp$ models the game in normal form as a two‑player noncooperative strategic game. Here,
	$\mathcal{N} = \lbr 1, 2\rbr$ represents the trustor and the trustee, respectively. The trustor's
	strategy set is $S_1 = \lbr\text{I}, \text{N}, \text{R}\rbr$, while the trustee's strategy set is
	$S_2 = \lbr\text{T}, \text{U}\rbr$, resulting in an overall strategy space $S = S_1 \times S_2$.
	The payoffs are given by a valued function $u: S \to \mathbb R^2$, where $u\lp s_1, s_2\rp = \lp u_1\lp s_1, s_2\rp, u_2\lp s_1, s_2\rp\rp, s_1\in S_1, s_2\in S_2$.
	
	With this setup, the previous globally dominant strategy
	of the trustee, namely that of not returning any fraction
	of the investment, disappears. In fact, when the trustor
	adopts the R~strategy, the comparison of payoffs between
	returning and not returning depends on both~$r$ and~$\beta$,
	so that no single strategy is favored for any choice of
	the game parameters. Similarly, for the trustor, not investing
	is no longer the preferred strategy, and, in addition, investing
	is at least not worse than paying for a reward. Notably,
	however, the introduction of a reward strategy only affects
	nontrivial structures, whereas it has no impact on the collapse
	of trust in a well-mixed population (see Appendix \ref{AppA} for details).
	
	\subsection{Evolutionary dynamics on lattices}
	When the game interactions take place on a network,
	the separation of payoff calculation and strategy imitation
	can cause even a nondominant strategy to be evolutionarily
	stable.
	
	In many practical domains, interactions between trustors
	and trustees can be represented as a two-dimensional(2D) grid structure
	with local connections, consisting of $L$~rows and $L$~columns
	of nodes with periodic boundary conditions. To distinguish
	between the two roles, we can categorize them according
	to the parity of the lattice coordinates. Thus, each node~$i$
	has von Neumann neighbors (up, down, left, right), denoted
	as $\Omega_i^+$, that play a different role, and diagonal
	neighbors (topleft, bottomleft, topright, bottomright),
	denoted as $\Omega_i^\times$, that have the same role as~$i$.
	
	As a result, the game interactions of node~$i$, which determine
	its payoff, occur with its $\Omega_i^+$ neighbors. Conversely,
	the strategy of~$i$ is determined by imitating its $\Omega_i^\times$
	neighbors. This is similar to the cross-learning framework of
	Ref.~\cite{wang2025inter}, with the difference that here the payoff
	interactions are inter-role, whereas the strategy learning is intrarole,
	as illustrated in Figs.~\ref{fig:network}(b) and~\ref{fig:network}(c).
	
	Using the payoff matrix in Eq.~(\ref{eq:payoff_matrix}),
	we define the actual payoff~$\pi_i$ of each node as the
	average payoff from all pairwise games with its $\Omega_i^+$
	neighbors. In formulas,
	\begin{equation}\label{eq:average_payoff}
		\pi_i = \frac{1}{\left|\Omega_i^+\right|}\sum_{k \in \Omega_i^+} u_{\operatorname{role}(i)}\lp s_i, s_k\rp\:,
	\end{equation}
	where $\operatorname{role}(i) = \ls\lp x_i + y_i\rp \mod 2\rs + 1$
	determines the role of node~$i$ and, consequently, its payoff function,
	from its lattice coordinates~$(x_i, y_i)$, and~$s_i$ and~$s_k$ are
	the strategies of node~$i$ and node~$k$, respectively. Note that,
	since we are modeling the game on a 2D lattice topology, the size
	of the interaction neighborhood is fixed, with $\left|\Omega_i^+\right| = 4$
	for all nodes.
	
	\begin{figure}
		\centering
		\includegraphics[width=0.55\textwidth]{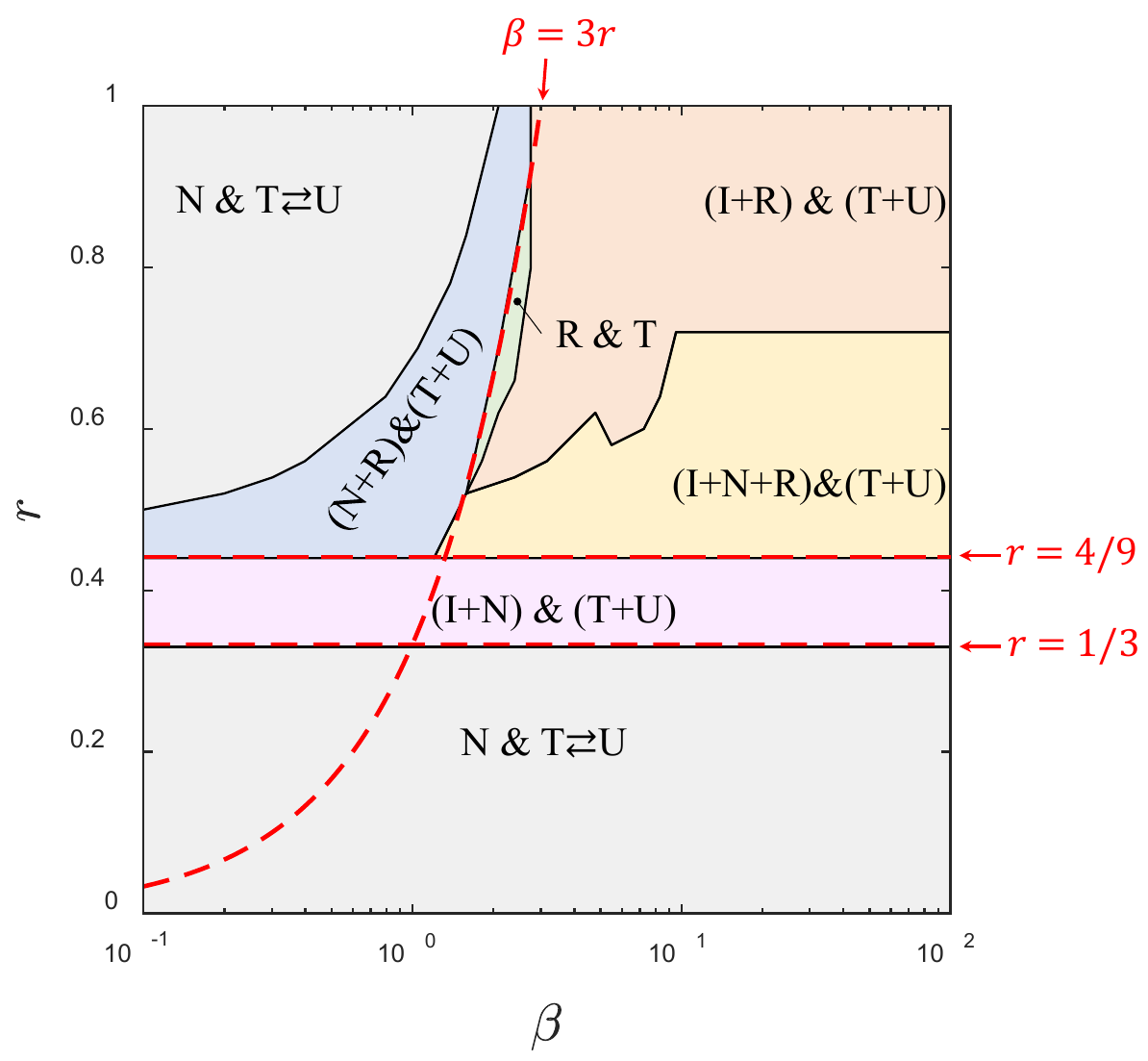}
		\caption{Trust mainly emerges when returns promote
			investments and rewards offset the cost of reciprocation. The $(\beta, r)$ phase diagram for fixed $\gamma = 0.1$ shows
			that the system admits trust as a stable strategy when the
			return ratio~$r$ is moderate, consistently with the conclusions
			of the classic two-strategy trust game. However, unlike what
			happens in the classic setting of the game, in which large
			return ratios suppress trust, our reward mechanism can support
			it for large enough rewards. The regions of different colors
			represent the stable strategy populations found for the corresponding
			parameters. Note that when the only strategy of the trustors
			is not to invest~(N), the two strategies of the trustees are
			completely equivalent.}\label{fig:beta_r_phase}
	\end{figure}
	The strategy evolution follows an asynchronous Fermi update rule.
	At each time step, a node~$f$ is selected uniformly at random from
	the network. A partner node~$m$ is then chosen from the learning
	neighborhood of~$f$, $\Omega_f^\times$. Their average payoffs~$\pi_f$
	and~$\pi_m$ are computed using Eq.~(\ref{eq:average_payoff}), and subsequently node~$f$ adopts the strategy
	of node~$m$ with a probability given by the Fermi function,
	\begin{equation}\label{eq:fermi}
		P_{f\leftarrow m}=\frac{1}{1+\e^{\lp\pi_f-\pi_m\rp/\kappa}}\:,
	\end{equation}
	where $\kappa$ = 0.1 is a noise parameter. 
	
	To study the behavior of our model, we carried out each simulation experiment on a $300\times 300$ grid. Each experiment involved 10 000~Monte~Carlo sweeps. A single Monte Carlo sweep comprised $L^2$ time steps [with one asynchronous update between a node pair occurring at each time step using Eq.~(\ref{eq:fermi})]. In each simulation run, the first 5 000 sweeps were used for system equilibration and subsequently discarded. The average ratio of trustor to trustee strategies was then computed over the subsequent 5 000 sweeps.
	
	\section{Results}
	
	According to our model definition, the system employs five strategies: I (trustor, invest), R (trustor, reward), N (trustor, not invest), T (trustworthy trustee), and U (untrustworthy trustee). Their respective proportions within each role are denoted as $\rho_\text{I}$, $\rho_\text{R}$, $\rho_\text{N} = 1 - \rho_\text{I}-\rho_\text{R}$, $\rho_\text{T}$, and $\rho_\text{U} = 1 - \rho_\text{T}$. To characterize the reciprocity within the population, we define the level of trust as the proportion of I and R, on the premise that the proportion of T is positive, i.e., $\rho_\text{I} + \rho_\text{R} \ \text{and}\ \rho_\text{T} > 0$~\cite{berg1995trust}. Since the invest strategy and reward strategy are inherently altruistic, their very existence implies that some trustors will invest funds, and trustees have sufficient motivation to return a portion of those funds. Thus, trust has been generated.
	
	Unless stated otherwise, the system is initialized on a square lattice with a bipartite topology [Fig.~\ref{fig:network}(b)]. While agent roles are spatially fixed, their initial strategies are assigned stochastically following a uniform distribution within each subpopulation (ensuring equal initial probability for all competing strategies). On a periodic lattice network of $L=300$, a population size of 90 000 agents is sufficiently large to ensure global convergence. For each parameter set, we conduct a simulation spanning 10 000 Monte Carlo sweeps (MCSs). Due to the large population size, independent runs yield identical results, making a single run representative (see Appendix \ref{AppB} for details).
	
	\subsection{Inter-role rewards can promote trust}
	To explore the role of rewards on the emergence of stable strategies,
	we first build the phase diagram of the system in the $(\beta, r)$ plane
	for fixed reward cost $\gamma=0.1$.
	
	The diagram, reported in Fig.~\ref{fig:beta_r_phase}, shows that, especially
	for rewards roughly smaller than~3, trust emerges as a stable strategy only
	when the return ratio~$r$ is moderate. This is consistent with the general conclusions
	of the classic two-strategy trust game, in which trust requires $r>1/3$ to emerge,
	but excessively high values of~$r$ eventually suppress it~\cite{wang2025inter}.
	The underlying mechanism can be described as follows: First, the return fraction
	must be at least~$1/3$ for the trustor to have any incentive to invest. But
	then, the trustee has a net advantage of~$3r$ (the payoff of strategy U minus the payoff of strategy T) when not reciprocating. Thus,
	as~$r$ increases, the~U strategy is increasingly profitable, thereby undermining
	the sustainability of investment.
	
	Conversely, for fixed values of~$r$ greater than~$4/9$, the mechanism induces the reemergence of trust as the reward value~$\beta$
	increases. In fact, in regions where both~$\beta$ and~$r$ are high,
	the noninvesting strategy~N completely disappears from the stable
	population, driven by the incentives offered to the trustees. This
	can be understood by noting that, when the trustor adopts the reward
	strategy, the trustee ends up with a profit of~$3(1-r)+\beta$ by
	reciprocating, but only~3 by defecting. Therefore, reciprocation
	becomes advantageous when $3(1-r)+\beta > 3$, i.e., when $\beta > 3r$.
	Only then does the reward sufficiently offset the cost of reciprocation,
	motivating the trustee to adopt the~T strategy. Thus, both for fixed~$\beta$
	and for fixed~$r$, the trust primarily emerges in the parameter
	regions satisfying $r > 1/3$ and $\beta > 3r$.
	
	\begin{figure*}[htbp]
		\centering
		\includegraphics[width=0.9\textwidth]{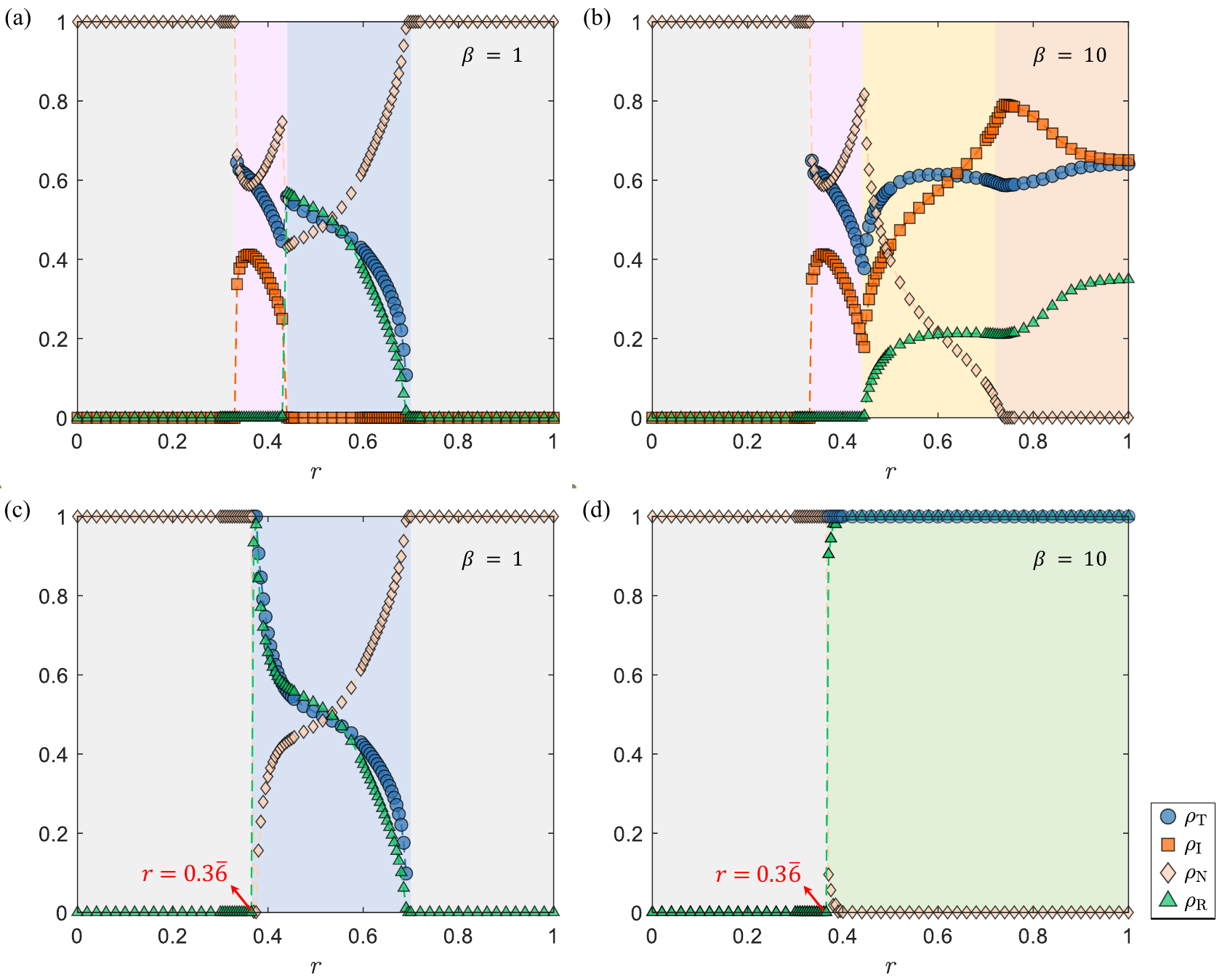}
		\caption{A sufficiently high reward can promote
			trust even at high returns. (a) If $\beta=1$, the reward
			value is insufficient, as the two inequalities $\beta > 3r$
			and $r > 1/3$ cannot be satisfied simultaneously, stopping
			trustees from adopting the reciprocation strategy~T, and
			thus not promoting the emergence of trust as~$r$ increases.
			(b) If $\beta=10$, the reward is sufficiently strong, satisfying
			$\beta > 3r$ for all values of~$r$. Thus, even in high-$r$
			regions, the mechanism drives trustees to reciprocate and
			trustors to choose investment or reward strategies, thereby
			promoting and stabilizing trust. (c) Except for removing strategy I, all other settings remain consistent with panel (a). Without strategy I suppressing strategy R, the reward strategy emerges at $r>0.3\overline{6}$. Subsequently, due to $\beta<3r$, the trust level decreases as r increases. (d) Except for excluding strategy I, all other settings remain consistent with panel (b). The reward strategy also emerges at $r > 0.3 \overline{6}$. Due to the reciprocal condition $\beta > 3r$, the population subsequently remains in the $\text{R}\&\text{T}$ phase regardless of how $r$ varies. In both panels, the reward
			cost for trustors is fixed at $\gamma=0.1$, and the color
			code is the same as in Fig.~\ref{fig:beta_r_phase}.}\label{fig:scan_r}
	\end{figure*}
	To better characterize the evolution of trust for different
	reward values, we study how the fractions of strategies at
	steady state change as a function of the return ratio~$r$ at
	$\beta=0.1$ and $\beta=10$. The results, reported in Figs.~\ref{fig:scan_r}(a) and \ref{fig:scan_r}(b),
	shed light on the individual regions of the phase diagram.
	
	For $\beta=1$, under the trust emergence condition $r > 1/3$,
	the inequality $\beta>3r$ never holds. Thus, as~$r$ increases,
	the trustees lose any incentive to choose the T~strategy. Eventually,
	the trustors completely abandon the I~strategy and split between
	noninvesting and offering a reward. While this causes a discrete
	jump in the adoption of trust, the reward mechanism is ultimately
	too weak, and the fraction of trustors choosing the R~strategy
	monotonically decreases with~$r$, eventually leading to the complete
	suppression of investment.
	
	In contrast, for $\beta=10$, the inequality $\beta > 3r$
	always holds for any value of~$r$, so that, so long as trustors
	invest, the reward makes the payoff for trustees choosing~T
	greater than that for trustees choosing~U. Then, as $r$~increases,
	the returns obtained by trustors improve, while trustees
	still maintain a relative advantage after receiving rewards.
	These factors synergistically drive an increase in the proportion
	of~I and the adoption of~R, strengthening trust overall.
	In turn, this causes a decrease of noninvestors~(N), which
	completely disappear at $r\approx 0.74$, and the stabilization
	of a strong majority of trustworthy trustees.
	
	\begin{figure*}[htbp]
		\centering
		\includegraphics[width=1\linewidth]{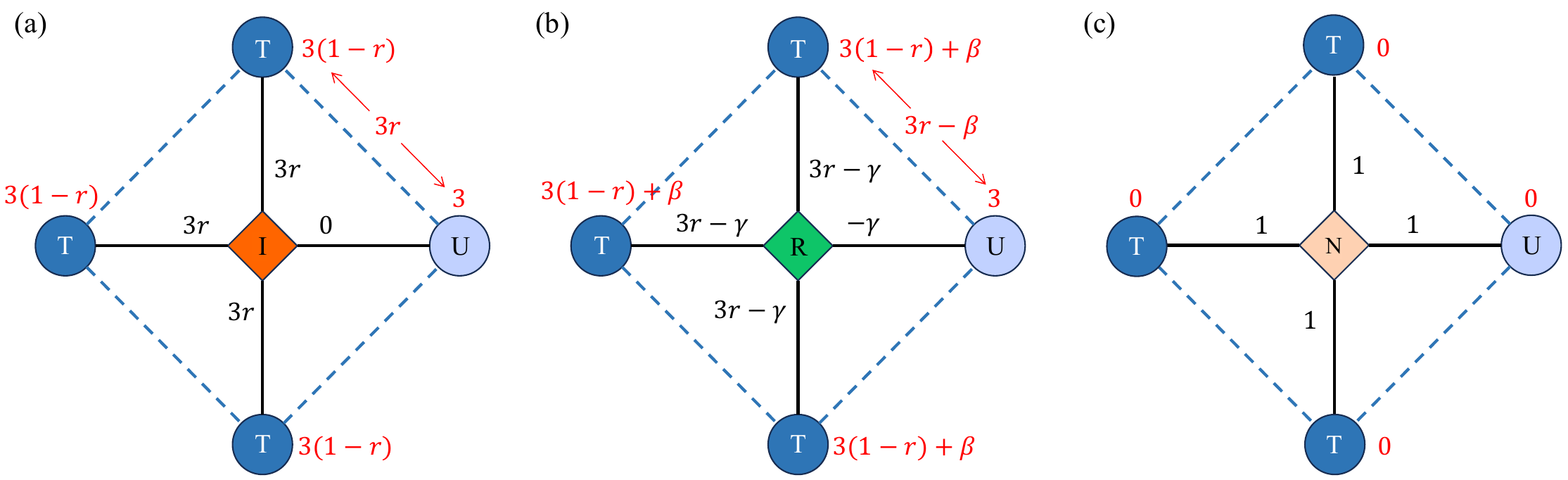}
		\caption{The sensitivity of I to U's invasion is crucial for maintaining I's survival, while the reward benefit of R is key to determining U's transformation into T. (a) When there is one U strategy among the trustee neighbors surrounding I, I's payoff is $\pi_I = 9r/4$. The payoffs for a single T or U are $3(1-r)$ or $3$, respectively (with a net payoff gap of $3r$). (b) When there is one U strategy among the trustee neighbors surrounding R, R's payoff is $\pi_I = (9r-4\gamma)/4$. The payoffs for a single T or U are $3(1-r)+\beta$ or 3, respectively (the magnitude of $\beta$ relative to $3f$ determines the conversion direction between T and U). (c) Regardless of the strategy distribution among the trustor's neighbors, the payoff for N is always 1, with a zero payoff difference between T and U among its neighbors.}
		\label{fig:U_invasion}
	\end{figure*}
	
	It is worth noting that regardless of how $\beta$ varies, $r=4/9$ serves as the threshold for the emergence of the R strategy, as evident from Figs.~\ref{fig:beta_r_phase},~\ref{fig:scan_r}(a), and~\ref{fig:scan_r}(b). In fact, $r=4/9$ represents the threshold for the extinction of the I strategy. When $r>4/9$, the I strategy becomes insensitive to U strategy invasions (failing to immediately convert to N). Comparing Figs~\ref{fig:U_invasion}(a) and~\ref{fig:U_invasion}(c), when a trustor's T-strategy environment is invaded by just one U strategy, if the payoff $\pi_\text{I}=9r$ from choosing the I strategy exceeds the payoff $\pi_\text{N}=4$ from the N strategy (i.e., $r$ is excessively large: $r>4/9$), the trustor will ignore the U invasion and persist with the I strategy. At this point, the payoff difference between T and U strategies surrounding the trustor is $3r$, which will eventually drive all neighbors to switch to the U strategy. Once all T strategies in the vicinity are eliminated, the I strategy will also perish. When strategy I dies out, its suppression of strategy R ceases. As shown in Fig.~\ref{fig:U_invasion}(b), the payoff difference between T and U surrounding R is $3r-\beta$. This reduced payoff difference delays U's invasion of T, favoring R's survival. Thus, when $r>4/9$, strategy R emerges.
	
	To better elucidate the suppression effect of I on R, we removed the I strategy from the population of trustors. We set the initial distribution conditions to $\rho_\text{R}, \rho_\text{T}\approx 50\%$. This yielded the proportion changes of the R, N, and T strategies as a function of $r$, as shown in Figs.~\ref{fig:scan_r}(c) and~\ref{fig:scan_r}(d).
	
	We find that the condition for the emergence of strategy R in panels (c) and (d) is $3r-\gamma>1$ (i.e., $r > (1 + \gamma)/3 = 0.3\overline{6}$), whereas this condition does not hold in panels (a) and (b). This is because the critical value $0.3\overline{6}$ still falls within the emergence range of strategy I, $1/3 \leq r \leq 4/9$, where strategy R is suppressed. The distinction between Figs. 3(a) and (c) lies in the interval $1/3 \leq r \leq 4/9$ where strategy I is present. Upon the extinction of strategy I, strategy R begins to exert its influence: It narrows the payoff gap between neighboring T and U, and even causes T's payoff to exceed U's ($\beta>3r$), as illustrated in Fig.~\ref{fig:U_invasion}(b).
	
	It can be observed that Figs.~\ref{fig:scan_r}(a) and~\ref{fig:scan_r}(b) exhibit a discrete jump and a relatively continuous change, respectively, at $r>4/9$. This primarily depends on whether $\beta<3r$ or $\beta>3r$. When $\beta < 3r$ [Figure \ref{fig:scan_r}(a)], the I strategy and R strategy play similar roles, both relying on the rapid transition of U invaders (I/R → N) to maintain trust. Therefore, the I strategy curve in Fig.~\ref{fig:scan_r}(a) resembles the R strategy curve in Fig.~\ref{fig:scan_r}(c). When I disappears, R emerges, exhibiting a discrete jump. However, when $\beta>3r$ [Figure~\ref{fig:scan_r}(b)], the emergence of the R strategy causes surrounding U to tend toward T. Thus, when $r>4/9$, the survival logic of the I strategy shifts from sensitivity to U to reliance on R. Once the I strategy ceases to suppress R, the emergence of R introduces more T individuals into the population, thereby stimulating the growth of the I strategy again. Since the environment consistently favors the generation of the T strategy, the proportion of R changes relatively slowly.

	\subsection{Inter-role rewards not always promote trust}
\begin{figure}
		\centering
		\includegraphics[width=0.45\textwidth]{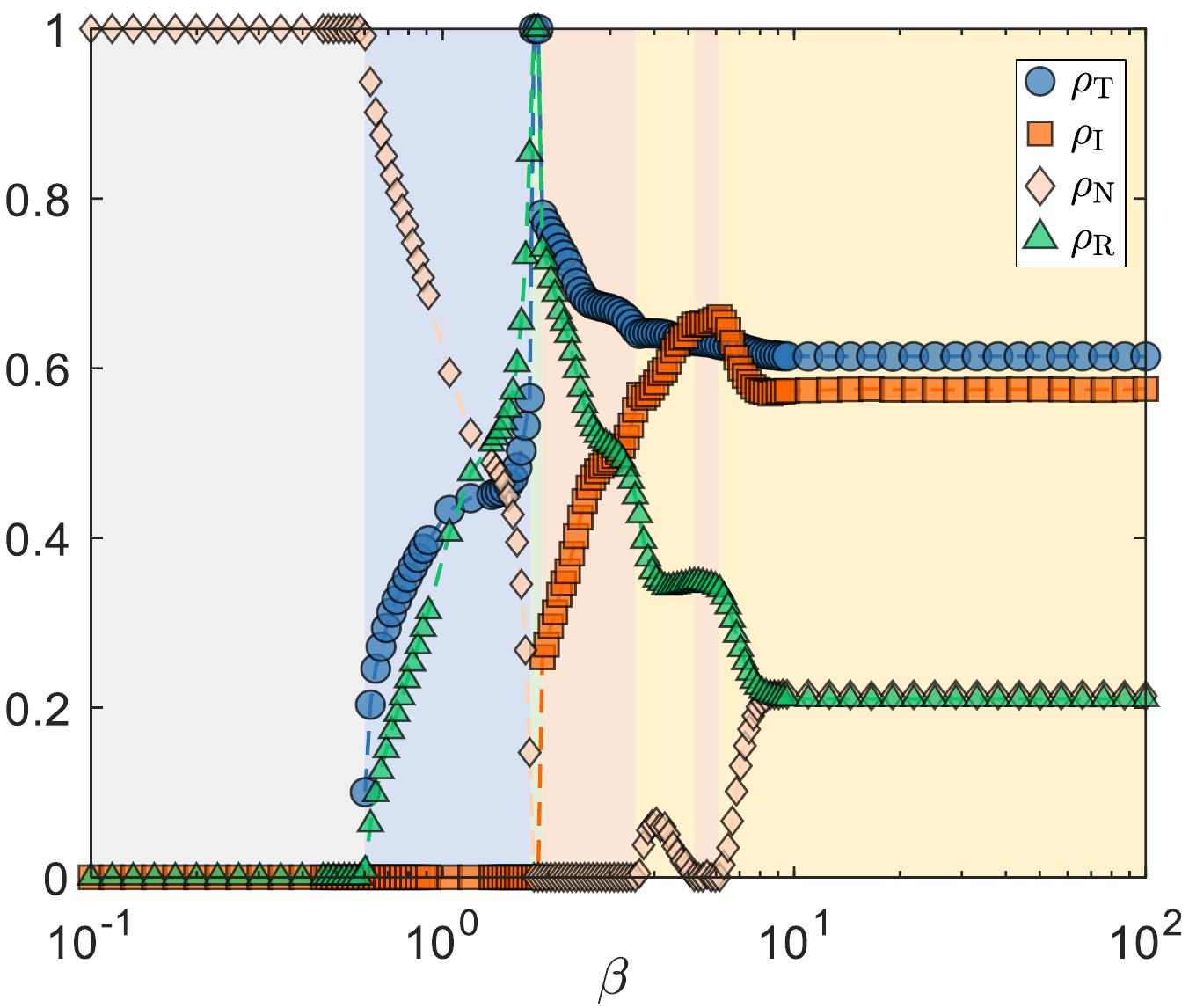}
		\caption{Greater rewards do not promote
			trust if the return ratios are not sufficiently
			large. For $\gamma=0.1$ and $r=0.6$, the adoption
			of reciprocation by the trustees reaches its maximum
			for $\beta=2$, decreasing afterward.}\label{fig:scan_beta}
	\end{figure}

	To more clearly characterize the impact of the reward value~$\beta$
	on the emergence of trust, we fix the return ratio at $r=0.6$, which
	crosses the most phase boundaries in the $(\beta,r)$ phase diagram,
	as seen in Fig.~\ref{fig:beta_r_phase}. Then, we vary~$\beta$, measuring
	the steady-state fractions of each strategy. The results, reported
	in Fig.~\ref{fig:scan_beta}, show that the proportion of I and R~strategy
	does not increase monotonically with the reward value. Rather, under
	the moderate return ratio chosen, trust is maximized by a moderate
	reward value, becoming the only chosen strategy at $\beta=2$, when
	the noninvestors disappear and the trustors only adopt the reward strategy. Conversely, the fraction of trustors choosing to invest is
	highest at $\beta=6$, where the noninvestors are still absent and
	the fraction of reciprocating trustees is approximately~$0.63$. However, when $\beta>6$, as $\beta$ increases, some trustors start to choose the noninvestment strategy, leading to a decline in the trust level.
	
	An analysis of the core mechanisms of the rewarded trust model explains
	why excessively large rewards can counterintuitively hinder the emergence
	of trust. From the point of view of the trustors,
	\begin{itemize}
		\item the I~strategy strictly dominates~R across all parameter regimes,
		as players invariably have an incentive to select~I over~R, especially
		as the reward cost~$\gamma$ increases;
		\item the preference between~I and~N hinges on whether trustees
		reciprocate, so that trustors shift from~N to~I only when trustees
		select~T and $3r > 1$; otherwise, if the trustees choose~U or $3r\leqslant 1$,
		N dominates~I;
		\item trustors prefer~R to~N only when trustees choose~T and $3r - \gamma > 1$;
		otherwise, if trustees adopt~U or $3r - \gamma \leqslant 1$,   N dominates~R.
	\end{itemize}
	For what concerns the trustees, instead,
	\begin{itemize}
		\item when trustors opt for~N, there is no difference   between choosing~T or~U, as the payoff vanishes anyway;
		\item when trustors select~I, trustees shift from~T to~U;
		\item when trustors adopt~R, trustees transition from~U   to~T only if $\beta > 3r$; otherwise, if $\beta \leqslant 3r$,   U strictly dominates~T.
	\end{itemize}
	These relations are visually summarized in Fig.~\ref{fig:trans}, where dashed lines denote transitions that require meeting their respective thresholds.
	
	These considerations make the mechanism underlying the suppression
	of trust at large values of~$\beta$ evident: Increasing $\beta$ increases
	the relative payoff of adopting a trustworthy strategy only when the
	trustor offers a reward, but at the same time it does not provide
	any direct benefit to the trustor. Concurrently, for the trustors, if the strategy distribution of surrounding trustees is identical,
	I is always preferable to~R, so the cluster of nodes adopting~R that
	may be temporarily induced by a higher~$\beta$ is ultimately replaced
	by~I under selective pressure. Subsequently, neighboring trustees
	revert from~T to~U due to the absence of rewards, thereby undermining
	the reciprocity framework. The net result is that large values of~$\beta$
	counterintuitively amplify a cascading sequence, causing trust to
	be lower than at moderate values.
	
	The results discussed above are qualitatively robust with respect
	to changes in the reward cost~$\gamma$. To show this, we computed
	the phase diagram of the system in the $(\beta, r)$ plane for $\gamma=0.5$.
	The results, illustrated in Fig.~\ref{fig:Stability}, show that,
	while the stable adoption of the reward strategy now requires a higher
	return fraction, slightly greater than~$0.5$, the region where trust
	reaches the theoretical maximum (green) becomes much larger than
	at $\gamma=0.1$. This indicates that the reward cost itself plays
	a complex role in determining the stable states of the system.
	
	Furthermore, Fig.~\ref{fig:Stability} indicates that the emergence conditions for strategies I and R are $1/3 < r < 4/9$ and $r > 0.5$, respectively. Notably, the threshold for the emergence of R ($3r - \gamma > 1$, yielding $r > 0.5$) falls strictly outside the survival interval of strategy I. Consequently, the emergence of R is decoupled from the competitive suppression of I. This leads to a distinct sequential transition as $r$ increases: Strategy I first emerges and then vanishes; following an intermediate phase dominated by strategy N, strategy R emerges. This reappearance fosters the selection of T by trustees, ultimately promoting the evolution of trust.
	
	\subsection{Low costs do not always favor the offer of a reward}
	To evaluate the effect of~$\gamma$ in more detail, we examine
	the $(\beta,\gamma)$ phase diagram of the system at $r=0.5$,
	reported in Fig.~\ref{fig:beta_gamma_scanGamma}. The clear division
	of the diagram into an upper part with no investment and a complex
	lower part shows that trust emerges only when $\gamma<0.5$, confirming
	the accuracy of the criterion $\gamma<3r-1$, which we discussed
	in the previous subsection. This indicates that, to effectively
	facilitate trusting relationships, reward costs cannot be excessively
	high. Similarly, a further division of the diagram can be observed,
	with trust emerging only when $\beta>3r$, consistently with our
	previous analysis.
	
	However, in the parameter region where both criteria are satisfied
	(bottomright in Fig.~\ref{fig:beta_gamma_scanGamma}), a counterintuitive
	behavior can be observed. In principle, one would expect that a lower
	reward cost would promote the adoption of a rewarding strategy on
	the part of the trustors, but the opposite effect is found, namely
	that the R~strategy is maximum in the area closest to the threshold
	$\gamma=0.5$, and its adoption actually decreases if the cost becomes
	lower.
	
	\begin{figure}
		\centering
		\includegraphics[width=0.65\textwidth]{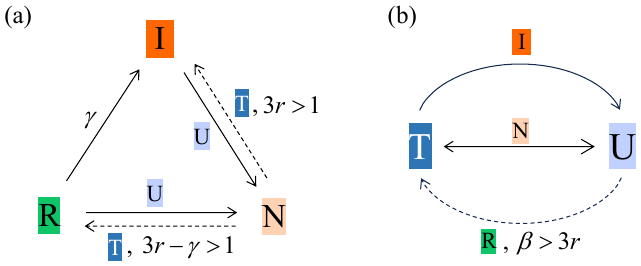}
		\caption{Strategy transition relationships.
			(a) Transition pathways among the three trustor strategies~I,
			N, and~R as influenced by trustee behavior. (b) Transition pathways
			between the two trustee strategies T and~U as influenced by trustor
			behavior. In both panels, arrows indicate shifts toward strategies
			yielding higher payoffs. Solid lines denote unconditional transitions,
			whereas dashed lines represent conditional transitions that require
			specific threshold criteria to be satisfied.}\label{fig:trans}
	\end{figure}
	To explain this observation, consider the mesoscopic evolutionary
	rules summarized in Fig.~\ref{fig:trans}. Assuming that the reward
	is greater than~$3r$, a very low value of~$\gamma$ will initially
	result in trustors that have adopted the N~strategy switching in
	roughly equal fractions to~R and~I. In turn, the trustees that interact
	with investors offering no reward are subject to pressure to switch
	from~T to~U. But then, nonreciprocating trustees tend to go back
	to trustworthy behavior when offered a reward, because of its advantageous
	value. This results in the red region in the phase diagram, where
	the trustors always invest and sometimes offer a reward, while the
	trustees partially return the fixed fraction of the investment.
	
	As the reward cost increases, however, the rate at which investing
	trustors stop offering rewards while keeping investing increases,
	while that of noninvesting trustors that switch to offering a reward
	decreases. The net effect of this is the appearance of~N as a stable
	strategy, resulting in the total coexistence phase (yellow). Note
	that, in this phase, there is a dynamic equilibrium between~N and~I,
	as well as one between~N and~R, because the inequality $3r>1$ is satisfied.
	
	When $\gamma$ increases even further, the faster rate at which trustors
	switch from~R to~I causes a transient adoption of the U~strategy by a large
	number of trustees, which, in turn, converts the trustors from~I to~N.
	However, the effective lack of~I causes a progressive depletion of the
	U~strategy among the trustees, which shifts the balance between~N and~R
	in favor of the latter. Ultimately, this nucleates a growing phase in
	which all the trustors offer a reward, and all the trustees return the
	investment, resulting in the phase where both the offer of a reward and
	trust are maximum.

	\section{Discussion}
	\begin{figure}
		\centering
		\includegraphics[width=0.55\textwidth]{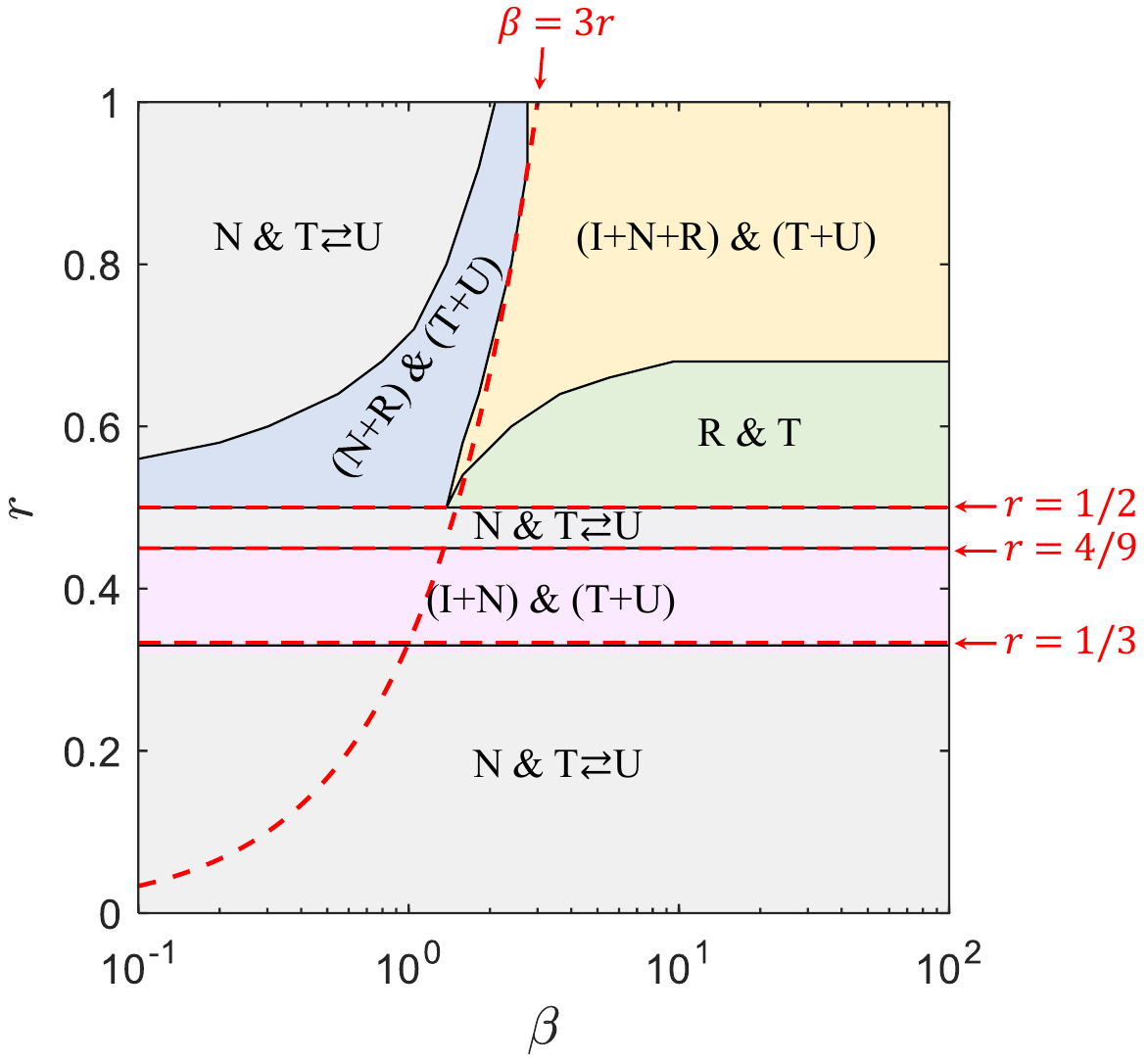}
		\caption{The features of the emergence of trust
			are stable with respect to the reward cost. The $(\beta,r)$
			phase diagram for $\gamma=0.5$ remains broadly the same
			as that for $\gamma=0.1$, illustrated in Fig.~\ref{fig:beta_r_phase}.
			The main difference is that the four-strategy coexistence
			phase, shown in red in Fig.~\ref{fig:beta_r_phase}, is
			replaced by the yellow five-strategy coexistence phase,
			which, in turn, leaves the place to the complete-trust
			phase (green).}\label{fig:Stability}
	\end{figure}
	In summary, we have established a spatial trust game to model
	and analyze how the introduction of a reward mechanism promotes
	the emergence of trust in structured populations. Our findings
	show several counterintuitive phenomena. First, trust emerges
	only under moderate rewards. This observation can be explained
	by considering that the cyclical structure formed by the three
	strategies of the trustors co-evolves with the strategies of
	the trustees. Thus, increases in the reward eventually cause a
	disbalance between the trustors, who do not receive any direct
	benefit from them, and the trustees, who only profit if they
	choose a trustworthy strategy and actually receive a reward.
	Ultimately, the overall trust becomes lower for higher rewards
	than it is for moderate ones.
	
	Previous studies on trust games without reward mechanisms
	have shown that strong relations of trust are only found
	for moderate values of the return ratio, which has been recognized
	as the most critical factor influencing their emergence.
	In our model, we find that higher rewards can indeed promote
	trust under high return rates. However, this relationship
	is nonmonotonic, and the maximum trust occurs for moderate
	levels of both reward and cost, whereas excessively high
	rewards cause an increase of the number of noninvestors
	and nonreciprocating trustees, as does an excessive reduction
	of the reward. Similarly, higher reward costs require correspondingly
	greater return rates or reward values to sustain the reward
	strategy.
	
	\begin{figure}
		\centering
		\includegraphics[width=0.55\textwidth]{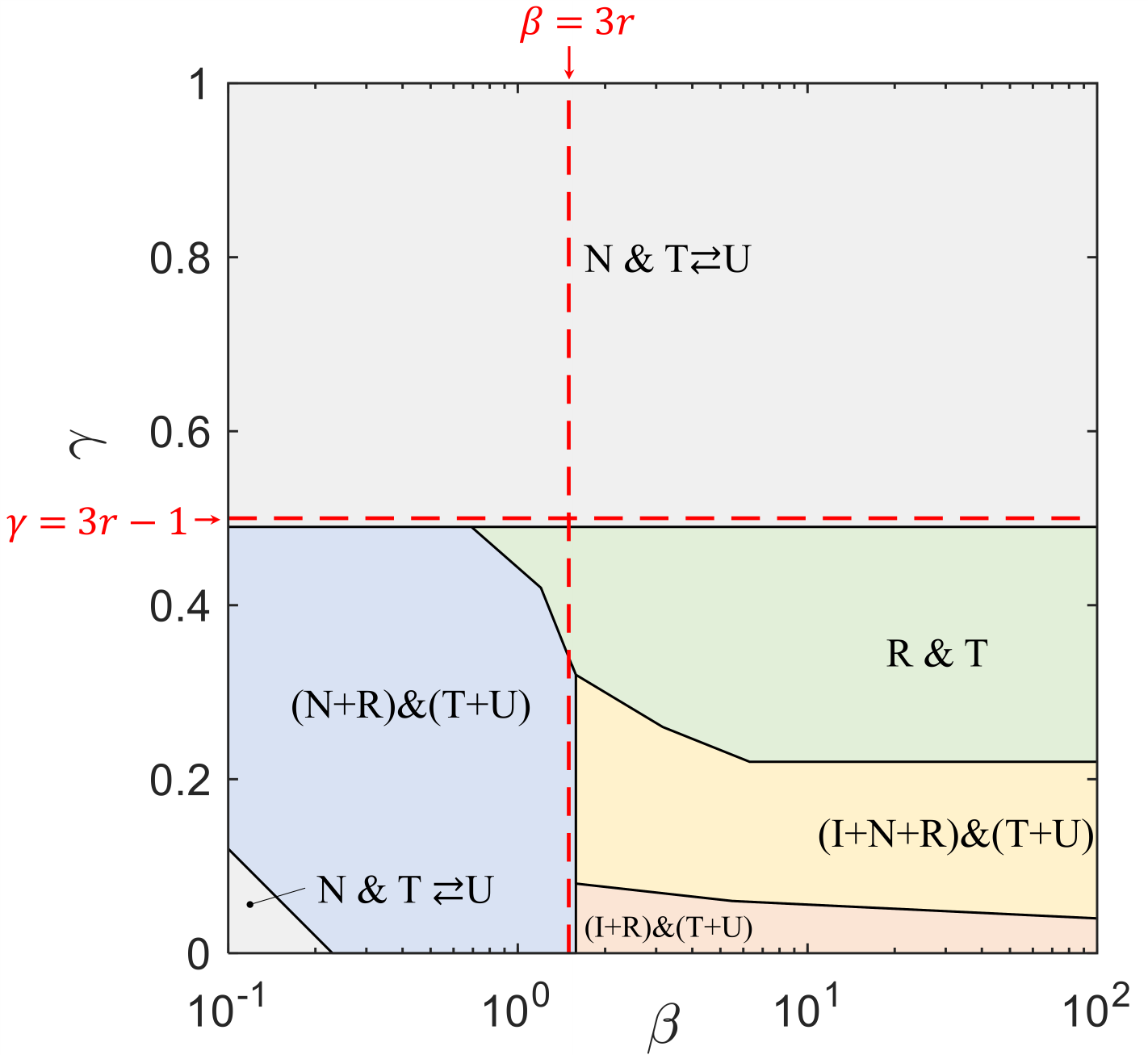}
		\caption{Reducing the cost of rewards does
			not always foster trust. The $(\beta,\gamma)$ phase
			diagram for fixed return ratio $r=0.5$ shows that
			the phase where the adoption of a reward strategy
			and trust are both maximum does not occur for the
			lowest reward cost~$\gamma$, but rather for intermediate
			values thereof, closer to the threshold $3r-1$.}\label{fig:beta_gamma_scanGamma}
	\end{figure}
	To explain how the coevolution mechanism of strategies
	causes the rich phase diagram observed, we carried out
	extensive numerical simulations of the model. These allowed
	us to describe how the system reaches each of its possible
	steady states, and to elucidate the interplay of the conversion
	rates between the N~strategy and the I~one, and the role
	played by the model parameters in balancing them. Given the latest theoretical advances in evolutionary game theory, future research may explore how the inherent asymmetry between trustors and trustees influences trust evolution, while also examining scenarios where network structures dynamically change over time. Incorporating spatial and temporal complexity information could enhance the robustness of the mechanisms identified in this study~\cite{su2022evolution_PNAS, su2023strategy}.
	
	In conclusion, by employing a minimal trust model augmented with a reward mechanism, we have uncovered complex emergent patterns in strategic transitions. Our work offers a framework for understanding
	multistrategy evolutionary dynamics, especially in the presence
	of complex transitional relationships, highlighting the subtle
	role of spatial structure in social dilemmas and providing insights into why reward mechanisms do not always promote
	trust in real-world scenarios. Additionally, our setup has
	the potential to be applied also to other complex systems,
	such as chemical reaction networks with catalytic processes
	involving multiple species or complex food webs in ecology,
	both of which exhibit dynamical changes in chemical or biological
	species composition.

	\appendix
	\section{Replicator equations in a well-mixed population}\label{AppA}
	\setcounter{equation}{0}

	Here, we analyze the behavior of our model in the thermodynamic limit and under the assumption of a well-mixed population. We denote the frequencies of trustor strategies as~$\rho_\text{I}$, $\rho_\text{N}$ and~$\rho_\text{R}$, and the frequencies of trustee strategies as~$\rho_\text{T}$ and~$\rho_\text{U}$.
	
	Based on the payoff matrix in Eq.~\ref{eq:payoff_matrix}, the expected payoffs for a trustor adopting each of the possible strategies are
	\begin{align}
		\bar\pi_\text{I} &= 3\rho_\text{T}r ,\\
		\bar\pi_\text{N} &= 1 ,\\
		\bar\pi_\text{R} &= 3\rho_\text{T}r-\gamma\:.
	\end{align}
	
	Similarly, the expected payoffs for a trustee adopting either of the avaliable strategies are
	\begin{align}
		\bar\pi_\text{T} &= 3\lp\rho_\text{I}+\rho_\text{R}\rp\lp 1-r\rp+\rho_\text{R}\beta ,\\
		\bar\pi_\text{U} &= 3\lp\rho_\text{I}+\rho_\text{R}\rp\:.
	\end{align}
	
	Then, let $\bar\pi^{[1]}=\rho_\text{I}\bar\pi_\text{I}+\rho_\text{N}\bar\pi_\text{N}+\rho_\text{R}\bar\pi_\text{R}$ and $\bar\pi^{[2]}=\rho_\text{T}\bar\pi_\text{T}+\rho_\text{U}\Bar\pi_\text{U}$ be the average payoffs of the trustor population and of the trustee population, respectively.
	
	Using replicator dynamics~\cite{taylor1978evolutionary}, the evolution of~$\rho_\text{I}$, $\rho_\text{R}$, and~$\rho_\text{T}$ can be expressed as
	\begin{align}
		\begin{split}
			\dot\rho_\text{I} &= \rho_\text{I}\lp\bar\pi_\text{I}-\bar\pi^{[1]}\rp \\
			&= \rho_\text{I} \ls 3\lp 1-\rho_\text{I}-\rho_\text{R}\rp\rho_\text{T} r-1+\rho_\text{I}+\rho_\text{R}\lp 1+\gamma\rp\rs
		\end{split} ,\\
		\begin{split}
			\dot\rho_\text{R} &= \rho_\text{R}\lp\bar\pi_\text{R}-\bar\pi^{[1]}\rp \\
			&= \rho_\text{R} \ls 3\lp 1-\rho_\text{I}-\rho_\text{R}\rp\rho_\text{T} r-1+\rho_\text{I}+\rho_\text{R}\lp 1+\gamma\rp-\gamma\rs
		\end{split} ,\\
		\begin{split}
			\dot\rho_\text{T} &= \rho_\text{T}\lp\bar\pi_\text{T}-\bar\pi^{[2]}\rp \\
			&= \rho_\text{T} \lp 1-\rho_\text{T}\rp\ls -3\lp\rho_\text{I}+\rho_\text{R}\rp r+\rho_\text{R} \beta\rs\:.
		\end{split}
	\end{align}
	
	From the equations above, it follows that the system has no interior equilibrium or isolated boundary equilibria. However, $\rho_\text{N} = 1$ forms a stable line $(0,0,\rho_\text{T}^\circ)$. On this line, $\dot\rho_\text{T} = 0$, and the value of $\rho_\text{T}$ is subject to neutral drift. The Jacobian matrix of this stable line is
	\begin{equation}
		J = \begin{pmatrix}
			3\rho_\text{T}^\circ r-1 & 0 & 0 \\
			0 & 3\rho_\text{T}^\circ r-1-\gamma & 0 \\
			-3\rho_\text{T}^\circ\lp 1-\rho_\text{T}^\circ\rp & \rho_\text{T}^\circ\lp 1-\rho_\text{T}^\circ\rp\lp -3r+\beta\rp & 0
		\end{pmatrix}\:,
	\end{equation}
	whose eigenvalues are $\lambda_1 = 3\rho_T^\circ r - 1$, $\lambda_2 = 3\rho_T^\circ r - 1 - \gamma$, and $\lambda_3 = 0$. When $3\rho_T^\circ r < 1$, it is $\lambda_1 < 0$ and $\lambda_2 < 0$, and the system evolves toward this stable line, so that, for a large population, the mean-field solution is that all trustors adopt the noninvesting strategy, and trust collapses.
	
	\section{Evidence of convergence across mulitple runs}\label{AppB}
\begin{figure}[htbp]
		\centering
		\includegraphics[width=0.68\linewidth]{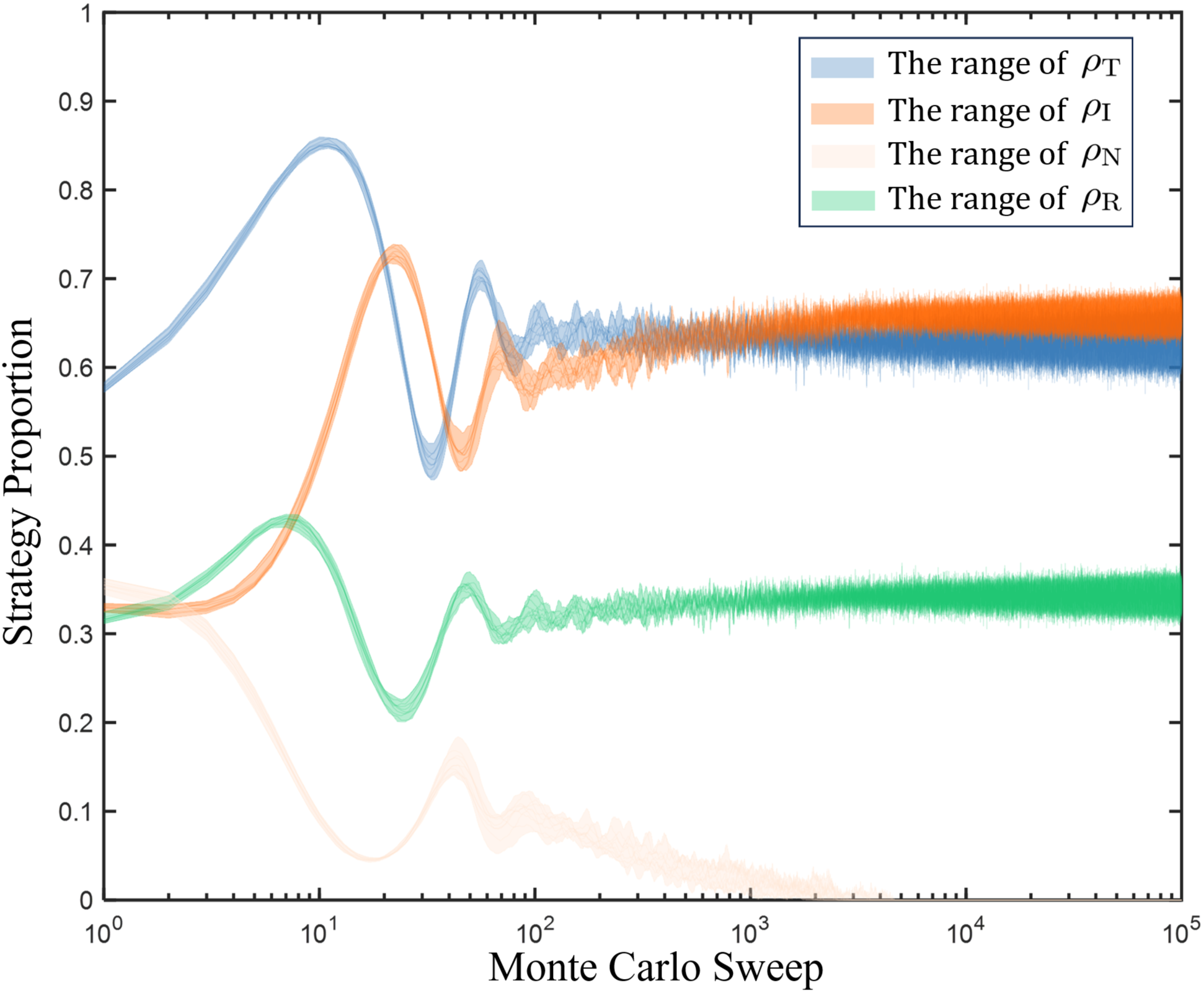}
		\caption{The strategy proportion within the population as it evolves through Monte Carlo sweeps. Each curve in the figure represents the temporal variation in strategy proportions formed during a single Monte Carlo run, with shaded areas indicating the range of variation across ten runs.} 
		\label{fig:TemporalLine}
	\end{figure}

	To verify the convergence of the simulation, we performed ten independent Monte Carlo runs with the parameter set $r=0.6, \beta=6, \gamma=0.1$, which corresponds to a representative data point in Figs.~\ref{fig:beta_r_phase} and~\ref{fig:scan_beta}. Figure~\ref{fig:TemporalLine} illustrates the temporal evolution of the fraction of each strategy within the population. To validate that the standard 10 000 MCSs are sufficient for the system to reach a steady state, we extended the simulation duration to 100 000 MCSs for this analysis.
	
	The results from ten independent Monte Carlo runs exhibit strong convergence throughout the entire evolutionary process. This consistency demonstrates that for large populations, the simulation outcomes are robust and representative. Consequently, the results presented in the figures are derived from a single representative simulation run.
	
	\begin{acknowledgments}
		This work was supported by the National Science and Technology Major Project for Brain Science and Brainlike Intelligence Technology (2025ZD0215700), the National Natural Science Foundation of China (72421002, 72401287, 72401289), the Hunan Provincial Department of Science and Technology (2025JJ60447), and the Major Program of Xiangjiang Laboratory
		(24XJJCYJ01001). C.I.d.G. acknowledges funding from the Bulgarian Ministry of Education and Science (BG-RRP-2.004-0006-C02). The authors declare that they have no conflict of interest.
	\end{acknowledgments}
	
	
	%

\end{document}